\input amstex
\documentstyle{amsppt}
\refstyle{C}

\NoBlackBoxes
\TagsOnRight
\CenteredTagsOnSplits

\magnification=1200
\hcorrection{-0.0625 in}
\vcorrection{0.0 in}
\pagewidth{5.6 in}
\pageheight{7.5 in}
\nopagenumbers
\loadbold

\topmatter

\title
Lebowitz Inequalities for Ashkin--Teller Systems
\endtitle

\leftheadtext { L. Chayes and K. Shtengel}
\rightheadtext\nofrills {Lebowitz Inequalities for AT Systems}

\author
\hbox{\hsize=2.75in
\vtop{\centerline{L. Chayes}
\centerline{{\it Department of
Mathematics}}
\centerline{{\it UCLA}}}
\vtop{\centerline{K. Shtengel}
\centerline{{\it Department of Physics}}
\centerline{{\it UCLA}}}}
\endauthor

\address
L. Chayes
\hfill\newline
Department of Mathematics
\hfill\newline
University of California
\hfill\newline
Los Angeles, California 90095-1555
\endaddress
\email
lchayes\@math.ucla.edu
\endemail

\address
K. Shtengel
\hfill\newline
Department of Physics
\hfill\newline
University of California
\hfill\newline
Los Angeles, California 90095-1547
\endaddress
\email
shtengel\@physics.ucla.edu
\endemail


\keywords
Lebowitz Inequalities, Ashkin--Teller Models, Random cluster and loop
representations
\endkeywords

\thanks
Work supported in part by a grant from the
NSA (MDA904-98-1-0518) and the NFS (99-71016)
\endthanks

\abstract
We consider the Ashkin--Teller model with negative four-spin  coupling
but still in the region where the ground state is ferromagnetic.  We
establish the  standard Lebowitz inequality as well as the extension
that is necessary to prove a divergent susceptibility.
\endabstract
\dedicatory
\medskip
Dedicated to Joel Lebowitz who has, throughout the years, maintained
the cohesion of the Statistical Mechanics community.
\enddedicatory
\endtopmatter
\newpage
\document

\vfill

\newpage
\pageno = 1

\TagsOnRight
\CenteredTagsOnSplits

\hcorrection{-0.1 in}
\vcorrection{-0.25 in}
\pagewidth{5.8 in}
\pageheight{8.0 in}

\baselineskip = 13pt

\subheading
{1. Introduction}
The Lebowitz inequality was first established for ferromagnetic
Ising  systems in \cite{1}.  It was later shown \cite{2,3}
that in these and  related systems such an inequality implies a
continuous transition. The inequality  itself states that the
(untruncated) four point correlation function is bounded above by
the sum of the products of the two point functions paired in all
possible ways.  From this one  can show that the derivative of the
susceptibility
\footnote{Throughout this work, we will use the word
``susceptibility'' to denote the sum of the {\it untruncated\/}
2-point function.  For the systems under  consideration, this
coincides with the thermodynamic susceptibility in the single phase
regime and it is the divergence of {\it this} object that we wish to
study.}
is bounded above by a constant times the square of the
susceptibility. And from this one can show that if at some $\beta^*$
the  susceptibility, $\Cal X$, is infinite, then for any
$\beta < \beta^*$ we have
$\Cal X \geq [\text{const.}][\beta^* - \beta]^{-1}$.
This  necessarily implies that the susceptibility has
to diverge continuously at some point.
\footnote{This is easily seen on a formal level in infinite volume;
a  rigorous proof often involves finite-volume cutoffs.  Such
derivations are routine  and will not be made explicit in this note.
For example the methods used in the percolation version of
these arguments \cite{4}, see also \cite{5} 
(with suitable modifications for boundary conditions  if
necessary) can be applied to every (uniform) system known to the
authors.}

In spin systems, the derivations of these sorts of inequalities
usually follow the same course:  Duplicate the system, rearrange 
variables, expand out all terms and hope everything comes out positive. 
(Although in \cite{1}, some FKG properties of the duplicated system were
also exploited.)  The model we  consider, introduced in
\cite{6}, is described by the Hamiltonian
$$
-\Cal H = \sum_{\langle i,j\rangle}
\left[J(\sigma_i\sigma_j + \tau_i\tau_j) 
- U\sigma_i\sigma_j\tau_i\tau_j \right]
\tag 1.1
$$
where $\sigma$ and $\tau$ represent Ising variables and $J$ and $U$
are both positive.
Although a
proof along the above described lines may be possible, it would be
extraordinarily complicated due to the presence of the negative
couplings.  (Not to mention the number of variables that would be
required.)  Our strategy -- more akin to the derivation in \cite{1} --
involves two distinct  steps.  First:  get a bound on the pure four-point
function (involving only the
$\sigma$-variables) in terms of the various mixed four-point functions.
Second:  use the  repulsive nature of the interaction between the two
types of variables to bound the mixed  four-point functions by products
of the corresponding two-point functions.  In short, no  duplication.  In
point of fact, we will present these steps in the reverse order.  In
Section 2, we describe a random cluster representation for this system
(introduced in \cite{7,8}) that provides the
bounds described in the second step.  In Section 3 we present a variant of
the so-called loop expansion which gets us  through the first step.
We remark, without understanding, that both representations have
their advantages and limitations.  In particular, a unified proof
using  either representation alone seems to be completely hopeless.
We remark without proof that  this general strategy can be
implemented in other systems as well, e\.g\. the  usual O(2) and O(3)
ferromagnets.  However the system described by the Hamiltonian in
Eq.(1.1) is the first {\it new\/} system for which a Lebowitz
inequality can be proved by  these methods.

It is clear that the behavior of the system depends drastically on the
magnitude of $U$.  E\.g\. if $U < J$, the ground state is ferromagnetic
and if $U > J$, it is not.  We will work in the ferromagnetic domain
which, it seems, is determined by the condition
$$
\text{th}^2 \beta J \geq  \text{th}\,\beta U.
\tag 1.2
$$

The fact that systems described by Equations (1.1) and (1.2) have a
continuous  transition is, perhaps, not too surprising.  Nevertheless,
there is only a small  handful of models for which a continuous
transition can be rigorously established. Now it is  slightly larger
handful.

\subheading {2. The Random Cluster Representation}
To implement the random cluster representation, we rewrite the
Hamiltonian in Potts form:

$$
-\beta\Cal H = \sum_{\langle i,j\rangle}
\left[D\left(\delta_{\sigma_i\sigma_j} + \delta_{\tau_i\tau_j}\right)
 - V\delta_{\sigma_i\sigma_j}\delta_{\tau_i\tau_j}\right]
\tag 2.1
$$
with $D = 2\beta(J + U)$ and $V = 4\beta U$.  For each bond, we may
expand as in the usual random  cluster model, e.g\.
$\exp\left(D \delta_{\sigma_i\sigma_j}\right) = \left(1 +
R_D\delta_{\sigma_i\sigma_j}\right)$, etc\. with $R_D = \roman{e}^D -1 $
but when we get to the four-spin term, we are forced  to acknowledge
that $\exp\left(-V\delta_{\sigma_i\sigma_j}\delta_{\tau_i\tau_j}\right) 
= \left(1 - r_V\delta_{\sigma_i\sigma_j}\delta_{\tau_i\tau_j}\right)$ 
with $-r_V = \left(\roman{e}^{-V} - 1\right)$ a negative number.  
This could ostensibly  lead
to negative numbers in the graphical representation -- and hence no
representation with a {\it probabilistic\/} interpretation.  However
the situation is not  nearly as bad as it looks since, on each bond,
all we are interested in is the product of  all three terms. Indeed,
it is seen that there are there are several mechanisms for generating
a $\delta_{\sigma_i\sigma_j}\delta_{\tau_i\tau_j}$ term and we get

$$
\exp{\left\{D\left(\delta_{\sigma_i\sigma_j} + 
\delta_{\tau_i\tau_j}\right) -
V\delta_{\sigma_i\sigma_j}\delta_{\tau_i\tau_j}\right\}} =
1 + R_D\delta_{\sigma_i\sigma_j} + R_D\delta_{\tau_i\tau_j}
+ \goth R_{VD}\delta_{\sigma_i\sigma_j}\delta_{\tau_i\tau_j};
\tag 2.2
$$
with $\goth R_{VD} = \left[R_D^2 - r_V( 1 + R_D)^2\right]$.  
Thus, to get
started on a decent graphical representation all we need is that
$\goth R_{VD} \geq 0$; i.e\. 
$\left(1 - \roman{e}^{-D}\right)^2  \geq 1 -\roman{e}^{-V}$.  
After some small amount of work,  this is seen to
be precisely the condition (1.2) mentioned in the introduction.

There is an obvious way to develop a graphical representation.
Following closely the derivation in \cite{9} for the Potts model, we
could define the $R_D\delta_{\sigma_i\sigma_j}$ and
$R_D\delta_{\tau_i\tau_j}$ terms as single bonds in the
$\sigma$-layer and $\tau$-layer respectively while the $\goth
R_{VD}$-terms represent  double bonds.  Then all the different types
of bonds are treated as separate entities.  This was done in
\cite{7} and is well suited for certain purposes but
none that are related to the present work.  An alternative approach,
used in both \cite{8} and \cite{7}, is to define bond
configurations $\Omega_{\sigma}$ and $\Omega_{\tau}$ for the $\sigma$
and $\tau$ layers that do not care where the bonds came from.  I.e\.
$\Omega_{\sigma}(b) = 1$ means that $b$ got a single $\sigma$-bond or
a maybe a double bond.  In this approach, the  resulting random
cluster measure is defined by the weights

$$
W(\Omega_{\sigma},\Omega_{\tau}) =
A^{[N(\Omega_{\sigma}\vee \Omega_{\tau})]}
B^{[N(\Omega_{\sigma}\wedge \Omega_{\tau})]}
2^{C(\Omega_{\sigma})}2^{C(\Omega_{\tau})}
\tag 2.3
$$
with $A = R_D$ and $B = \goth R_{VD}/R_D$.  As usual:  $N(-)$ is the
number of bonds of the specified type while $C(\Omega_{\sigma})$ and
$C(\Omega_{\tau})$ individually count the  number of connected
components in the $\sigma$-layer and $\tau$-layers respectively.
(Of course the latter objects -- as well as possible additional
constraints -- are not  completely defined until one pays heed to the
{\it boundary conditions}.)

It is also convenient to write the weights in the form where each
occupied bond gets its own factor of $A$ and then an additional factor
of $B/A$ for the number  of overlapping bonds. Then, as is is clear,
if $B > A$, then the bonds in the separate  layers attract while if $B
< A$, they repel.  And notice that the dividing line is determined  by
the sign of $U$. What has impeded progress so far is the
(straightforward) result concerning the FKG-properties of these
measures that was proved in  \cite{7,8} (see
also \cite{10}).  For future purposes, we state the result in general
terms (c.f\. the remark following Proposition 2.1)
\proclaim{Proposition 2.1}
Let $\Cal G$ denote a finite graph.  For $r, s \geq 1$
let $\mu_{\Bbb P}^{\text{RC}}(-)$ be the measure on pairs of bond
configurations
$(\Omega_\sigma, \Omega_\tau)$ with weights given by
$$
W_{\Bbb P}(\Omega_\sigma,\Omega_\tau)  =
r^{C(\Omega_\sigma)}s^{C(\Omega_\tau)}
\prod_b [A_\sigma(b)]^{\Omega_\sigma(b)}[A_\tau(b)]^{\Omega_\tau(b)}
\lambda_b^{[\Omega_\sigma(b)][\Omega_\tau(b)]}
$$
where $\Omega_\sigma(b) = 0$ or $1$ corresponds to the $\sigma$-bond
being vacant or occupied and similarly for $\Omega_\tau(b)$ and $\Bbb
P$ is notation for the various parameters in the weights.  Then the
sufficient -- and the necessary -- condition for the measure
$\mu_{\Bbb P}^{\text{RC}}(-)$  to be strong FKG is that $\lambda_b
\geq 1$ for all $b$.
\endproclaim
\demo{Proof}
This is a reasonably straightforward FKG lattice condition to check;
see \cite{7,8} (and also \cite{10}).
\qed
\enddemo

\remark{Remark}
The definition of the measures -- for the benefit of Propositions 2.1
and 2.2 -- is a little unwieldy.  The case in hand is simply
$A_\sigma(b) = A_\tau(b)  \equiv A$ and $\lambda_b \equiv B/A$.
(Thus the theorem fails for models of  interest in this paper; we will
return to this point after we finish apologizing for the  statement of
the proposition.) The primary reason we use ``the most general graphs''
is as follows: to complete the argument that gets us from a Lebowitz
inequality to a divergent susceptibility we need to consider these
measures in the context of sensible graphs -- like  finite pieces of
$\Bbb Z^2$ -- in the presence of boundary conditions e.g\. ``wired''.  On
a  given finite graph most boundary conditions that lead to FKG measures
for random cluster  problems can be realized as free boundary conditions
on a related graph.

As just mentioned:  with the usual ordering, the FKG property for the
systems described by the Hamiltonian in Eq.(2.1) will only hold in
case $-V  \geq 0$.  However, there is an alternative ordering which is
described in \cite{8}  that is well suited for the situation at
hand.  This alternative exploits the competitive  nature of the
interaction between the layers.  In particular, for any bond $b$, we
can define the local order
$\Omega_\sigma(b) = 1 > \Omega_\sigma(b) = 0$ and
$\Omega_\tau(b) = 0 > \Omega_\tau(b) = 1$. Notice that with this
ordering, there is no
relation between the configuration with
$(\Omega_\sigma(b) = 1,\Omega_\tau(b) = 1)$ and $(\Omega_\sigma(b) =
0,\Omega_\tau(b) = 0)$.
\endremark
\medskip
As obvious as all this may seem in hindsight, it is nevertheless
surprising:  If, for example, $U < J$, the low temperature states are
ordered in {\it both\/} layers:  Despite the competition, there is
ultimately some cooperation.  In any case, it is equally not  hard to
show that for $B < A$, the measures have the FKG property with respect
to this  alternative partial ordering.

\proclaim{Proposition 2.2}
Let $\mu_{\Bbb P}^{\text{RC}}(-)$ be a random cluster measure as
described in the statement of Proposition 2.1 with $\lambda_b \leq 1$
on each bond.  Then  $\mu_{\Bbb P}^{\text{RC}}(-)$ is FKG with respect
to the partial order where $\Omega_\sigma(b)$ and  $1 -
\Omega_\tau(b)$ are $($separate$)$ increasing coordinates.
\endproclaim
\demo{Proof}
See \cite{8} Lemma 4.1.
\qed
\enddemo

The significance of this representation for the present work is
exemplified by the following  (which already appears in \cite{8}):
\proclaim{Corollary}
For the spin systems described in Proposition 2.2 let $i$, $j$, $k$
and $\ell$ denote any sites of the lattice.  Then, for $r = s = 2$,
$$
\langle \sigma_i\sigma_j\tau_k\tau_\ell \rangle_{\beta\Cal H}
\leq
\langle \sigma_i\sigma_j \rangle_{\beta\Cal H}
\langle \tau_k\tau_\ell \rangle_{\beta\Cal H}.
$$
For the other spin systems of this type, the same holds with
$\sigma_i\sigma_j$ replaced by $\frac
r{r-1}\left(\delta_{\sigma_i\sigma_j} -\frac 1r\right)$, etc.
\endproclaim
\demo{Proof}
According to the standard translations between spin systems and their
random cluster representations (\cite{11} or, better yet, \cite{12})
the object 
$\langle \delta_{\sigma_i\sigma_j}  - \frac 1r \rangle_{\beta\Cal H}$  
is proportional to the probability that the
sites $i$ and $j$ belong to the same $\Omega_{\sigma}$-\thinspace
cluster. Let us define the various  connectivity events: $\big\{i
\underset \sigma \to \longleftrightarrow j \big\}$ is the event that $i$
and $j$ are connected in the $\sigma$-layer  etc. Then we have, for $r
= s = 2$
$$
\langle \sigma_i\sigma_j\tau_k\tau_\ell \rangle_{\beta\Cal H} =
\mu^{\text{RC}}_{\beta\Cal H}\Big(\big\{i \underset \sigma \to
\longleftrightarrow j \big\}\cap
\big\{k \underset \tau \to \longleftrightarrow \ell \big\}\Big)
\tag 2.5
$$
(and similarly for the other values of $r$ and $s$).  However
$$
\mu^{\text{RC}}_{\beta\Cal H}\Big(\big\{i \underset \sigma \to
\longleftrightarrow j \big\}\cap
\big\{k \underset \tau \to \longleftrightarrow \ell \big\}\Big) 
\leq \mu^{\text{RC}}_{\beta\Cal H}\Big(\big\{i \underset \sigma \to
\longleftrightarrow j \big\}\Big)
\mu^{\text{RC}}_{\beta\Cal H}\Big(\big\{k \underset \tau \to
\longleftrightarrow \ell \big\}\Big)
$$
by the FKG property and the desired result is established.
\qed
\enddemo
We will use the above inequality -- as well as some related
inequalities --  when the
time comes to derive our principal results.
\subheading {3. The Loop Representation}
We now start down an  entirely different track namely that of loop
expansions.  The primitive versions of these  expansions are well
known (e\.g\. they are described in \cite{8}) and, in the context of
the  present model, actually date back to \cite{6}.  However, the
usual versions will only take  us so far and some small modifications
will eventually have to be inserted.  The unfortunate  {\it general\/}
limitation is that these expansions are heavily tied with the Ising
nature of the spin variables  (or, more precisely the ability to
represent the spin  variables in terms of Ising variables).  Thus, for
the remainder of this paper, we will confine attention to the standard
($r = s =  2$) Ashkin--Teller models.

We start with the famous Ising--spin identity:
$\roman{e}^{\kappa\sigma_i\sigma_j} \propto 1 +
(\text{th}\,\kappa)\sigma_i\sigma_j$.
In any Ising-type system, this can be applied to any set that has an
Ising-type interaction. But we remark that schemes based on this
identity usually require the coefficients of the Ising spin terms to
be positive.  For the case at  hand we get
$$
\roman{e}^{\beta[J(\sigma_i\sigma_j + \tau_i\tau_j) -
U\sigma_i\sigma_j\tau_i\tau_j]} \propto (1 + \Bbb J\sigma_i\sigma_j)
(1 + \Bbb J\tau_i\tau_j) (1 -\Bbb U\sigma_i\sigma_j\tau_i\tau_j)
\equiv \Phi_{i,j}
\tag 3.1
$$
With $\Bbb J = \text{th}\,\beta J$ and $\Bbb U = \text{th}\,\beta U$.  
If we multiply these out, the coefficient of the four-spin term is 
seen to be proportional to  $\Bbb J^2 - \Bbb U$ which we want to be
positive.  This is, again, the condition (1.2). Multiplication of
all terms leads to a couple of two-spin terms, a four-spin term and a
constant (that gets rescaled to one).  The result is 
$$
\Phi_{i,j}  \propto  \left(1 + \Bbb L\sigma_i\sigma_j + \Bbb L\tau_i\tau_j
+ \Bbb K\sigma_i\sigma_j\tau_i\tau_j\right)
\tag 3.2
$$
with $\Bbb L =  [\Bbb J - \Bbb U]/[1 - \Bbb J^2\Bbb U]$ and $\Bbb K =
[\Bbb J^2 - \Bbb U]/[1 - \Bbb J^2\Bbb U]$.  The further  development
of the loop expansion is somewhat obscured by the presence of the
four-spin term  so to simplify our arguments, we will start with the
treatment of the case $\Bbb K = 0$  -- saturation of the condition
(1.2).  (As a matter of fact, this very model was  discussed in
\cite{13} where, for example, the $2d$-version was shown to have a {\it
random surface\/}  interpretation.)  We assume that we are on some finite
graph that has free boundary conditions.  Let us start  with the
partition function:  We get to pick one term from $\Phi_{i,j}$ for
each bond of  the lattice.  One such choice for each bond leads to a
graphical configuration (or diagram)  which we will call $\eta$: a red
bond at $\langle i,j \rangle$ if we choose the $\Bbb
L\tau_i\tau_j$-term, a  blue bond if we pick $\Bbb L\sigma_i\sigma_j$
and vacant for the ``$1$''.  When we perform  the trace (normalized
for convenience) the only diagrams that survive are those where an
even number -- possibly zero -- of bonds of {\it both\/} types are
incident at each vertex.  When these constraints are satisfied, the
weight of the configuration is just $\Bbb  L^{|\eta|}$.  It is of some
use to consider  $\Xi$, a colorless configuration, that represents
all  the configurations $\eta$ that have the same bonds regardless of
color.  Although it is  tempting proclaim that the weight of $\Xi$ is
given $\Bbb L^{|\Xi|}\times 2^{\#\text{of loops in }\Xi}$,  the number
of loops in a configuration is not very well defined -- certainly not
in any way  that makes this formula true.  (Notwithstanding, we will
still call this the loop representation.) In the end, the coefficient
of $\Bbb L^{|\Xi|}$ is just the number of ways that the loop clusters
can be consistently colored red and blue in such a way that the
constraints are satisfied  at each vertex.  The trivial, but important
observation is that this counting procedure  factors over the
individual clusters of $\Xi$.

The spin--spin correlation function is not particularly pretty in
this representation -- it is not the probability of anything in the
loop measure (i.e\. the  measure defined by the weights of the loop
configurations).
\footnote{In general the correlation functions can be expressed as
the expected value of some function of $\Xi$ -- but this does not seem
to serve any  tangible purpose.  For certain correlation functions we
can do better.  For example we will show that
$\langle \sigma_i\sigma_j\tau_i\tau_j \rangle$ is just the
probability that $i$ and $j$ belong to the same loop (c\.f\. Corollary
I to Theorem 3.4).  Notice  that if $\Bbb U = 0$ the former is equal
to $\langle \sigma_i\sigma_j\rangle^2$.  This is a disguised version
of the famous ``switching identities'' derived in \cite{14}.}
When we calculate the correlation between $\sigma_i$ and $\sigma_j$
the best description is to say that there is a numerator and a
denominator.  The denominator  is just the partition function -- the
sum of all the weights described above.  The numerator contains
similar sorts of configurations except that now the two sites $i$ and
$j$ have to have an {\it odd\/} number of incident (blue) bonds.
Since every other site  has an even number of bonds, this implies that
there is a walk from $i$ to $j$.  Following the language of \cite{14}
we say there are {\it sources\/} at the sites $i$ and $j$.

Let us now consider the four-point function
$\langle \sigma_i\sigma_j\sigma_k\sigma_\ell  \rangle$. Here we have
sources at $i$, $j$, $k$ and $\ell$.  When we look at the
contributing terms, we see that there are two distinct possibilities:
All four sites are part of the  same connected cluster (regardless of
color) or the sites pair up forming two distinct  clusters -- and
there are three ways of doing this.  We denote the sum of the
corresponding terms by  $\Gamma_{i,j,k,\ell}$ for the case where the
four points are tied up in the same cluster,  $\goth g_{[i,j\mid
k,\ell]}$ for $i$ and $j$ in a separate cluster from $k$ and $\ell$
etc.  Thus  we may write
$$
\Cal Z_{\Cal R} \times \langle \sigma_i\sigma_j\sigma_k\sigma_\ell
\rangle_{\Cal R}
= \left[\Gamma_{i,j,k,\ell}
+ \goth g_{[i,j\mid k,\ell]}
+ \goth g_{[i,k\mid j,\ell]}
+ \goth g_{[i,\ell\mid j,k]}\right]
\tag 3.3
$$
where $\Cal R$ is the interaction that has $\Bbb J^2 = \Bbb U$.

Now let us consider a mixed four-point function, say
$\langle \sigma_i\sigma_j\tau_k\tau_\ell \rangle_{\Cal R}$.
Several things are immediately clear:  First, there are no terms like
the last two for this correlation function -- we need the proper
sources to be paired.  Second, there {\it will\/} be terms like the
first two -- a term with all four points connected  and one with $(i\
\&\ j)$ and $(k\ \&\ \ell)$ separately paired.  Thus we may
tentatively write $\Cal Z_{\Cal R}\langle
\sigma_i\sigma_j\tau_k\tau_\ell \rangle_{\Cal R} =
\tilde\Gamma_{i,j,k,\ell} + \tilde\goth g_{[i,j\mid k,\ell]}$.  Third,
it is clear that $\tilde\goth g_{[i,j\mid k,\ell]}  =  \goth
g_{[i,j\mid k,\ell]}$.  Indeed in any colored graph that contributes
to $\goth g_{[i,j\mid k,\ell]}$, one can  simply reverse the colors of
the cluster of $(k\ \&\ \ell)$.  This shows that
$\tilde\goth g_{[i,j\mid k,\ell]}  \geq  \goth g_{[i,j\mid k,\ell]}$
and the reverse
argument shows
$\goth g_{[i,j\mid k,\ell]}  \geq  \tilde\goth g_{[i,j\mid k,\ell]}$.

What is not immediately obvious is that the four-point cluster terms
are equal.  This is the subject of the argument below.
\proclaim{Lemma 3.1}
Let $\boldsymbol \Gamma$ denote a finite connected graph that
contains among others four points $i, j, k$ and $\ell$.  All sites in
$\boldsymbol \Gamma$ are  evenly coordinated except the above
mentioned which are odd-coordinated.  Let  $N_{b(i,j,k,\ell)}^\Gamma$
denote the number of ways that the bonds of this graph can be colored
red  and blue such that an even number of red and blue bonds are
incident at each vertex save  for $i, j, k$ and $\ell$ which have an
odd number of blues.  Let $N_{b(i,j);r(k,\ell)}^\Gamma$  denote the
number of similar colorings except this time the sites $k$ and $\ell$
have an  odd number of reds.  Then
$$
   N_{b(i,j,k,\ell)}^\Gamma  =  N_{b(i,j);r(k,\ell)}^\Gamma.
$$
\endproclaim
\demo{Proof}
Let $\goth p:k\to\ell$ denote a self-avoiding path in $\boldsymbol
\Gamma$ with terminal points $k$ and $\ell$.  Let $\Cal
N_{b(i,j,k,\ell)}^{\thinspace  \Gamma}$ be the {\it set\/} of
successful colorings of $\boldsymbol \Gamma$ according to the first
set of rules (which is observed to be non-empty) and similarly for
$\Cal  N_{b(i,j);r(k,\ell)}^{\thinspace \Gamma}$.  We will use $\goth
p$ to define a map from $\Cal N_{b(i,j,k,\ell)}^\Gamma$ to $\Cal
N_{b(i,j);r(k,\ell)}^{\thinspace \Gamma}$ which is one-to-one; the
map is simply to reverse the color of each bond along the path.

It is clear that the application of this map to an element of $\Cal
N_{b(i,j,k,\ell)}^{\thinspace \Gamma}$ produces a coloring of  the
desired type: Indeed, at internal sites of the path, the change in the
number of  blues is $\pm 2$ or zero with a corresponding change in
reds of $\mp 2$ or zero.  Similarly at  the endpoints the number of
blues changes by $\pm 1$ accompanied by a change in the  number of
reds by $\mp 1$.

Next, consider two distinct colorings of $\boldsymbol \Gamma$ in $\Cal
N_{b(i,j,k,\ell)}^{\thinspace \Gamma}$.  Either the two  colorings
have some difference on the compliment of $\goth p$ -- in which case
the target  configurations are surely different, or they must be
different along the path itself  -- in which case they are still
different when their paths get reversed.

Hence this map is indeed one-to-one which establishes
$N_{b(i,j);r(k,\ell)}^\Gamma  \geq  N_{b(i,j,k,\ell)}^\Gamma$.
But  we can construct a similar sort of map from
$\Cal N_{b(i,j);r(k,\ell)}^{\thinspace \Gamma}$ to
$\Cal N_{b(i,j,k,\ell)}^{\thinspace \Gamma}$ which gives us the
reverse inequality.
\qed
\enddemo
As an immediate consequence we get that $\Gamma_{i,j,k,\ell} =
\tilde\Gamma_{i,j,k,\ell}$.  This leads to the following preliminary
result:
\proclaim{Theorem 3.2}
Consider any graph $\Cal G$ with free boundary conditions and
interaction
$$
-\Cal R  =  \beta\sum_{i,j}(\sigma_i\sigma_j + \tau_i\tau_j) -
\alpha\sum_{i,j}\sigma_i\sigma_j\tau_i\tau_j
$$
with $\roman{th}\,\alpha = \roman{th}^2\beta$ \it.  Then
$$
\langle \sigma_i\sigma_j\sigma_k\sigma_\ell \rangle_{\Cal R}  \leq
\langle \sigma_i\sigma_j\rangle_{\Cal R}\langle\sigma_k\sigma_\ell
\rangle_{\Cal R}
+  \langle \sigma_i\sigma_k\rangle_{\Cal R}\langle\sigma_j\sigma_\ell
\rangle_{\Cal R}
+  \langle \sigma_i\sigma_\ell\rangle_{\Cal R}\langle\sigma_j\sigma_k
\rangle_{\Cal R}.
$$
\endproclaim
\demo{Proof}
On the basis of what has transpired we have
$$
\langle \sigma_i\sigma_j\sigma_k\sigma_\ell \rangle_{\Cal R}  =
\langle \sigma_i\sigma_j\tau_k\tau_\ell \rangle_{\Cal R}
+  \langle \sigma_i\sigma_k\tau_j\tau_\ell \rangle_{\Cal R}
+  \langle \sigma_i\sigma_\ell\tau_j\tau_k \rangle_{\Cal R}
-2\frac{\Gamma_{i,j,k,\ell}}{\Cal Z_{\Cal R}}.
\tag 3.4
$$
So we neglect the final term and use the FKG property from Section 2.
\qed
\enddemo

Of course the above is of no immediate benefit for the analysis of
the susceptibility; in this model, there are additional terms when we
differentiate the two-point  correlation function with respect to the
temperature.  However, the extra ingredients required  are not
particularly difficult:
\proclaim{Theorem 3.3}
Consider the system described in the statement of Theorem 3.2.  Then
$$
\langle \sigma_i\sigma_j;
[\sigma_k\sigma_\ell + \tau_k\tau_\ell -
\alpha^\prime \sigma_k\sigma_\ell\tau_k\tau_\ell]\rangle_{\Cal R}
\leq
\langle \sigma_i\sigma_k\rangle_{\Cal R}\langle\sigma_j\sigma_\ell
\rangle_{\Cal R}
+  \langle \sigma_i\sigma_\ell\rangle_{\Cal R}\langle\sigma_j\sigma_k
\rangle_{\Cal R}.
$$
where, in general, $\langle A;B \rangle =
\langle AB \rangle - \langle A \rangle \langle B \rangle$ and
$\alpha^\prime = \roman{d}\alpha/ \roman{d}\beta$.
\endproclaim
\demo{Proof}
We first remark that all we really require is that the coefficient of
the four-spin term is not less than $-1$.  We observe that in the
present case, $\alpha^\prime = 2\,\text{th}\,\beta/[1 + \text{th}^2\beta]
\leq 1$.  In  light of Theorem 3.2, all that we need to show is that
$\langle \sigma_i\sigma_j; [\tau_k\tau_\ell - \alpha^\prime
\sigma_k\sigma_\ell\tau_k\tau_\ell]\rangle_{\Cal R}$ is not positive.
Using $\alpha^\prime \leq 1$, we start with $(1 - \alpha^\prime
)\langle \sigma_i\sigma_j; \tau_k\tau_\ell \rangle \leq 0$ by the FKG
inequality.  Now what is left is to show that
$\langle \sigma_i\sigma_j; [\tau_k\tau_\ell -
\sigma_k\sigma_\ell\tau_k\tau_\ell]\rangle_{\Cal R} \leq 0$. We go
back to the FK representation and consider the untruncated quantity
$\langle \sigma_i\sigma_j \tau_k\tau_\ell(1  -
\sigma_k\sigma_\ell)\rangle_{\Cal R}$.  It is clear that this vanishes
unless $k$ is connected to $\ell$ by $\tau$-bonds.  But this also
vanished unless $k$ is {\it not\/} connected to $\ell$ by
$\sigma$-bonds.  If both of  these are satisfied, {\it and\/} the site
$i$ is $\sigma$-connected to $j$, we get one.  There are other
possibilities for getting something non-zero but these all come out
negative.  We thus have
$$
\langle \sigma_i\sigma_j [\tau_k\tau_\ell -
\sigma_k\sigma_\ell\tau_k\tau_\ell]\rangle_{\Cal R} \leq
\mu^{\text{RC}}_{\Cal R}\Big(\big\{i \underset \sigma \to 
\longleftrightarrow j \big\}\cap
\big[\big\{k \underset \tau \to \longleftrightarrow \ell \big\}\cap
\big\{ k \underset \sigma \to 
{\gets\negthickspace\times\negthickspace\to} \ell\big\}\big]\Big)
\tag 3.5
$$
where $\{ k \underset \sigma \to
{\gets\negthickspace\times\negthickspace\to}\ell\}$
is notation for the the compliment of the event
$\{k \underset \sigma \to \longleftrightarrow \ell \}$.  The second
event on the right hand
side (the one in the square brackets) is clearly decreasing while the
first one is
increasing.  Thus
$$
\langle \sigma_i\sigma_j
[\tau_k\tau_\ell - \sigma_k\sigma_\ell\tau_k\tau_\ell]\rangle_{\Cal R}
\leq \mu^{\text{RC}}_{\Cal R}\Big(\big\{i \underset \sigma \to
\longleftrightarrow j \big\}\Big)\ \mu^{\text{RC}}_{\Cal R} 
\Big(\big\{k \underset \tau
\to \longleftrightarrow \ell \big\}\cap\big\{ k \underset \sigma 
\to {\gets\negthickspace\times\negthickspace\to} \ell\big\}\Big).
\tag 3.6
$$
The second factor on the right hand side of Eq.(3.6)  is seen to equal
$\langle \tau_k\tau_\ell[1 - \sigma_k\sigma_\ell] \rangle_{\Cal R}$
and the desired version
of the Lebowitz inequality is established.
\qed
\enddemo

The corollary to the above should be that the susceptibility  diverges.
However before such a claim can be made, it has to be verified that the
susceptibility  is actually infinite somewhere.  Unfortunately with the
above interaction, this can only  be established for the model on the
square lattice; here it seems likely that here the  critical value is in
fact $\beta = \infty$ (c\.f\. the discussion in \cite{13}).
\proclaim{Corollary}
For the model on $\Bbb Z^2$ with the interaction as described in the
statements of Theorem 3.2, the susceptibility diverges continuously at
some value --  possibly infinite -- of the parameter $\beta$.
\endproclaim
\demo{Proof}
Infinite susceptibility, at $\beta = \infty$, was established for
this model in \cite{13}
using duality, the rest follows from the bound on
$\roman{d}\Cal X/\roman{d}\beta$
that results from the
inequality in the statement of Theorem 3.3
\qed
\enddemo

As emphasized by the above limitations, it is clearly desirable to
extend this set of results past the $\Bbb K = 0$ restrictions.  Going
back to Eq\.(3.2)  -- with $\Bbb K > 0$ -- the straightforward
procedure would be to include one new type  of bond, namely a
red--blue (double) bond that has weight $\Bbb K$.  As indicated,
these bonds have double occupancy but all the other constraints are
pretty much the same as  before.  Although this is the standard way of
doing these loop expansions, it is clear  that our analysis will run
into trouble.

First, on the \ae sthetic level, there is now no worthwhile expansion
for the model in terms of colorless graphs.  Every double bond forces
two colors through  that bond.  This means that under certain
circumstances, the color of one bond in a cluster could  determine the
coloring scheme of the entire cluster -- even though the cluster
itself could  have many ``loops''. On a more disturbing note:
Although an expression like Eq\.(3.3) is  still valid for the usual
four-point function, and similarly for the mixed four-point
functions, there is no obvious correspondence between the various
$\Gamma$-terms.  (We remind the  reader that these are the terms
corresponding to the diagrams where all four points are in the  same
cluster.)  Indeed it is clear, after a moments reflection, that each
of these  $\Gamma$'s contain diagrams that are conspicuously absent in
the other.  (The relevant $\goth  g$-terms {\it do\/} coincide but
this is small consolation.)  The resolution of all of  these
difficulties is, perhaps, obvious in hindsight:  We will borrow from
the one.

What we actually do is as follows:  We write $1 = (1-\Bbb K) + \Bbb K$
(noting that $\Bbb K < 1$) so that
$$
\align \Phi_{i,j} & \propto \big([1 -
\Bbb K] + \Bbb J \sigma_i\sigma_j + \Bbb J\tau_i\tau_j + \Bbb
K\sigma_i\sigma_j\tau_i\tau_j + \Bbb K\big) \\ 
& \propto (1 + \bold{J} \sigma_i\sigma_j + \bold{J}\tau_i\tau_j +
\bold{K}\sigma_i\sigma_j\tau_i\tau_j + \bold{K}) 
\tag 3.7 \endalign
$$
with $\bold{K} = \Bbb K/[1 - \Bbb K]$ etc.  The ``new'' terms
corresponding to the selection of the $\bold{K}$ may, perhaps, be
called {\it dud-bonds}.  In the colored graphs these have no special
significance -- they do not change the  color parity of any vertex --
but in any colorless representation, they have the same  appearance,
not to mention the same weight, as the red--blue type of double bond.

This is the bottom line, anything else may be regarded as window
dressing. Notwithstanding, an interpretation that is pleasing (at
least to the  authors) is as follows:  Each edge of the lattice can
potentially be split into two edges, an ``upper'' edge and a ``lower''
edge.  Let us write the $\bold{K}$ terms in Eq.(3.7) as $\bold{K} =
\frac 12 \bold{K}^{\text{rr}} + \frac 12 \bold{K}^{\text{bb}}$ and
$\bold{K}\sigma_i\sigma_j\tau_i\tau_j = \frac 12
\bold{K}^{\text{rb}}\sigma_i\sigma_j\tau_i\tau_j + \frac 12
\bold{K}^{\text{br}}\sigma_i\sigma_j\tau_i\tau_j$ with (of course)
$\bold{K} = \bold{K}^{\text{bb}} = \bold{K}^{\text{rr}} =
\bold{K}^{\text{rb}} = \bold{K}^{\text{br}}$. Now, in a double bond
situation (which automatically implements the ``bond splitting''
option) one of these four terms is selected. The four choices are
identified as double blue, double red, upper--red/lower--blue and
upper--blue/lower--red i.e\. the two edges of the split actually
colored. The constraints are exactly as they were before: an even
number of colors of each type at each vertex.  With this
interpretation in mind, it is  clear that we once again have a
bonified -- and useful -- colorless  representation.  Explicitly, for
a given configuration, the graph becomes a graph with single  edges
and split edges.  The weight is [a factor of $\bold{J}$ for each
(occupied)  single bond and a factor of $\frac 12 \bold{K}$ for each
(occupied) double] $\times$  [the number of ways that the
configuration can be consistently colored].  Notice that in  any
coloring in which a particular double ends up red/blue, there is
another coloring  -- regarded as distinct -- with all other bonds the
same but the particular bond a blue/red. Similarly for exchanging a
red/red with a blue/blue (but notice that the mono-colored and
di-colored doubles cannot be exchanged without forcing other changes
in the configuration.)  Thus, when all is said and done,  we are
back to the situation we had at $\Bbb K = 0$ -- albeit on the peculiar
graph  defined by the splits.

As a consequence obtain, rather inexpensively,
\proclaim{Theorem 3.4}
Consider a finite graph with the Hamiltonian

$$
-\Cal H =  \sum_{\langle i,j \rangle}J_{i,j}[\sigma_i\sigma_j 
+ \tau_i \tau_j] - 
\sum_{\langle i,j \rangle}U_{i,j}[\sigma_i\sigma_j\tau_i \tau_j]
$$
with
\rm$ \text{th}^2\beta J_{i,j} \geq \text{th}\,\beta U_{i,j} \geq 0$ \it
on every edge $\langle i,j \rangle$.  Then
$$
\langle \sigma_i\sigma_j\sigma_k\sigma_\ell \rangle_{\beta\Cal H}
\leq \langle \sigma_i\sigma_j\rangle_{\beta\Cal
H}\langle\sigma_k\sigma_\ell \rangle_{\beta\Cal H}
+  \langle \sigma_i\sigma_k\rangle_{\beta\Cal
H}\langle\sigma_j\sigma_\ell \rangle_{\beta\Cal H}
+  \langle \sigma_i\sigma_\ell\rangle_{\beta\Cal
H}\langle\sigma_j\sigma_k \rangle_{\beta\Cal H}.
$$
and
$$
\langle \sigma_i\sigma_j;
[\tau_k\tau_\ell -
\sigma_k\sigma_\ell\tau_k\tau_\ell]\rangle_{\beta\Cal H}
\leq 0
$$
\endproclaim
\demo{Proof}
The second inequality is a direct consequence of the FKG property
(that was established in Proposition 2.2) by following the argument in
Theorem 3.3.  Thus we  are left with the extension of Theorem 3.2 to
non-zero $\Bbb K$. Let us start by repeating Eq.(3.3) and the similar
equation for the mixed four-point function without notational
changes.  Obviously $\tilde\goth g_{[i,j\mid k,\ell]}$ still equals
$\goth g_{[i,j\mid  k,\ell]}$ by blatant color-reversal symmetry. Of
more significance, we claim that with  the dud-bond representation,
(and the ensuing interpretation) the extension of Lemma 3.1 is
immediate.  In particular the argument that was used to show
$\Gamma_{i,j,k,\ell} = \tilde \Gamma_{i,j,k,\ell}$ (whereby some path
connecting $k$ to $\ell$ gets color reversed) still goes through --
the formulation was for a general underlying graph and thus works for
the graph defined with  the splits.  Given the identities
$\Gamma_{i,j,k,\ell} = \tilde \Gamma_{i,j,k,\ell}$ and
$\tilde\goth g_{[i,j\mid k,\ell]} = \goth g_{[i,j\mid k,\ell]}$
(as  well as the FKG property) the desired result is manifest.
\qed
\enddemo
As a bonus, we obtain a direct relationship between a correlation
function and a loop
probability:
\proclaim{Corollary I}
Consider the system as described in the statement of Theorem 3.4.  Let
$\Cal L_{i,j} = \Cal L_{i,j}(\beta\Cal H)$ denote the probability, in
the loop representation, that the sites $i$ and $j$ belong to the same
loop  cluster.  Then
$$
\Cal L_{i,j}  =
\langle \sigma_i\tau_i\sigma_j\tau_j \rangle_{\beta\Cal H}.
$$
\endproclaim
\demo{Proof}
The numerator of the right-hand side requires a blue walk {\it and\/}
a  red walk between the sites $i$ and $j$.  Applying the color
switching techniques  described in Lemma 3.1 in the context of the
dud-bond representation -- as was done in  Theorem 3.4 it is seen that
each such term is in one-to-one correspondence with a term
contributing to $\Cal L_{i,j}$ and vice versa.
\qed

\enddemo
\proclaim{Corollary II}
Consider a periodic graph on which there is Hamiltonian $\Cal H$
(also periodic) of the type described above with $U_{i,j} \equiv
\beta^{-1}\bold{A}_{i,j}(\beta)$ where the $\bold{A}_{i,j}(\beta)$ are
smooth functions satisfying \rm $\text{th}\,\bold{A}_{i,j}(\beta) \leq
\text{th}^2 \beta J_{i,j}$ \it as well as
$\bold{A}_{i,j}^\prime(\beta) \leq J_{i,j}$ and $\lim
_{\beta\to\infty} \beta^{-1}\bold{A}_{i,j}(\beta) < J_{i,j}$. Here
$\beta_0 < \beta < \infty$ with $\beta_0$ small enough to ensure that
at $\beta = \beta_0$, the  susceptibility is finite.  Then there is a
$\beta_c$ along this trajectory where the  susceptibility diverges
continuously; in particular at least as fast as a constant times
$(\beta_c - \beta)^{-1}$.
\endproclaim
\remark{Remark}
One can presume that under the stated conditions every ferromagnetic
transition in the region $U > 0$ is covered.  Although this may not
seem evident in the physical variables used above, it is quite
transparent in the  $A$--$B$ variables that were featured in Section
2.  For example, in the homogeneous case  with $J \equiv 1$ and
$\beta U \equiv \alpha$, let us set $B = \lambda A$ with $\lambda < 1$ and
allow $A$ to range in $[0, \infty)$.  Then we find that
$r_V = (1 - \lambda)[A/(1  + A)]^2$ and, in general, $\alpha^\prime =
r_V/[1 - r_V(1 + A)]$.  Thus $\alpha^\prime < 1$ because the worst
case is the one with the biggest value of $r_V$ (namely  $\lambda =
0$) which is the borderline case discussed in Theorem 3.3.  Now
starting at $A = 0$ there is exponential clustering for all $A$ small
enough.  But for $A$ large  and $\lambda > 0$ the graphical
representation discussed in Section 2 has  percolation (in both
layers) which implies magnetization.  (These facts can be verified by
simple-minded comparisons to independent percolation.)  Thus, along
these curves,  there is always a point where the susceptibility is
infinite.  So,  for all intents and  purposes, this region has a phase
boundary separating $\beta = 0$ from $\beta =  \infty$.  For negative
values of $B$ (i\.e\. $\lambda < 0$) the ground states are not
ferromagnetic and it seems highly unlikely that ferromagnetism could
somehow be generated by  entropic effects.
\endremark
\demo {Proof}
The above hypotheses are sufficient to derive bounds of the form
$\roman{d}\Cal X/\roman{d}\beta \leq \text{[const\.]}\Cal X^2$
(suitably regularized in  finite volume if necessary).  Furthermore,
they are sufficient to ensure a  ferromagnetic transition at finite
$\beta$ (c\.f\. the above remark) and hence $\Cal X$ is  actually
infinite somewhere.  The combination of these ingredients implies the
divergence of the susceptibility with the stated bound.
\qed
\enddemo

\enddocument

\end